\documentclass[a4paper,aps,pra,twocolumn,showpacs,preprintnumbers,amsmath,amssymb]{revtex4}
\usepackage{bbm}
\usepackage{amsmath}
\usepackage{verbatim}
\usepackage{graphicx}
\bibliography{References}
\begin{document}

\title{Detecting entanglement in two mode squeezed states by particle counting}

\author{Christine A. Muschik$^1$, Eugene S. Polzik$^{2,3}$ and J. Ignacio Cirac$^1$}

\affiliation{ $^1$Max-Planck--Institut f\"ur Quantenoptik,
Hans-Kopfermann-Strasse, D-85748 Garching, Germany \\
$^2$ QUANTOP, Danish Research Foundation Center for Quantum Optics, DK 2100 Copenhagen, Denmark\\
$^3$ Niels Bohr Institute, DK 2100 Copenhagen, Denmark}

\begin{abstract}
We present an entanglement criterion for two mode squeezed states
which relies on particle counting only. The proposed inequality is
optimal for the state under consideration and robust against
particle losses up to $2/3$. As it does not involve measurements of
quadratures - which is typically very challenging for atomic modes -
it renders the detection of atomic many-particle entanglement
feasible in many different settings. Moreover it bridges the gap
between entanglement verification for a qubit and criteria for
continuous variables measured by homodyne detection. We illustrate
its application in the context of superradiant light scattering from
Bose Einstein condensates by considering the creation of
entanglement between atoms and light as well as between two
condensates in different momentum states. The latter scheme takes
advantage of leaving the Gaussian realm and features probabilistic
entanglement distillation.
\end{abstract}

\pacs{}

\maketitle

\section{Introduction}
Entanglement is a true quantum feature. The study of this
peculiarity of  physics does not only hold the promise to acquire a
deeper understanding of Nature, but also paves the way towards
auspicious applications of quantum information science such as
unconditionally secure communication, ultraprecise measurements,
quantum computing and quantum simulation. Therefore, the quest for
entanglement or inseparability criteria is a vigorous field of
research \cite{EntanglementCriteria}.

In the most basic case, entanglement is shared between two parties
holding a single particle each and bipartite entanglement of single
pairs is well understood. Then again, entanglement between parties
holding a huge number of particles was studied for Gaussian states
with great success \cite{CVentanglement}. Here, the natural question
arises, how entanglement can be verified in the intermediate regime
and how the two well-studied cases cases of single pairs and
Gaussian modes can be linked by an inseparability criterion for an
arbitrary number of particles, which is not restricted to Gaussian
states. Starting from this motivation, we bridge the gap between
inseparability criteria for a qubit and continuous variables for
entanglement in two mode squeezed states. Moreover, from a practical
point of view, the new entanglement condition provides experimental
feasibility and applicability in numerous settings.

More specifically, entanglement in two mode squeezed states can, in
principle, be detected by means of a Gaussian inseparability
criterion \cite{EntQu}, which requires the measurement of variances
in canonical quadratures. This can be conveniently performed for
light modes, as first demonstrated in \cite{EPRlight}, as well as
for multi-atom collective spin modes, as shown in \cite{JKP}.
However, in many cases involving multi-atom entanglement homodyne
measurement of atomic canonical variables which require an atomic
"local oscillator" is not feasible.

This problem can be overcome by means of the practical entanglement
condition put forward in this work. The proposed inequality requires
only particle number measurements, rather than measurement of
quadratures and can be used to detect $N$-particle entangled states
of the form $|TMSS\rangle \otimes |TMSS\rangle$, which is attractive
in view of many recent experiments, that offer the potential to
generate this type of entanglement.

For example, superradiant scattering \cite{SR} of laser light from a
Bose-Einstein condensate was observed recently \cite{SRexp}.
Superradiant scattering leads to highly directional emission of
light from the atomic sample.  This striking effect attracted
considerable interest and it was shown theoretically that two mode
squeezed states can be generated in this context \cite{MM99, CP04,
Ent}. This is particularly interesting, since it represents the
interspecies atom-light analogue of photonic twin-beams generated in
optical parametric down conversion, which plays an essential role in
many applications of quantum optics and quantum information theory.
Despite the fact that the dynamics of the process and the resulting
non-separable state are well understood and the system is known to
be a very promising candidate for entanglement generation
\cite{MM99, CP04, Ent, EntProp}, entanglement could not be verified
in this system due to the absence of a suitable inseparability
criterion.
Thus, nonclassical correlations have been studied, but since the
quantum states produced in an experiment cannot be assumed to be
pure, correlations do not imply the presence of entanglement.
Other examples can be found in many different setups, as for example
in entanglement production in spin exchange collisions in
Bose-Einstein condensates \cite{DuanPu}, the free electron laser
\cite{BPRS06}, the creation of two mode squeezed states by
dissociation \cite{Diss} and entanglement production in colliding
Bose-Einstein condensates \cite{CorCol} or four-wave mixing in
matter waves in an optical lattice \cite{4WaveMixing}. While
correlations could be observed, multi-particle entanglement has not
yet been detected in this context. This gap can be closed by means
of the entanglement criterion put forward in this work.
It belongs to the class of entanglement conditions derivable from
the partial transposition criterion \cite{H.H.} and as the Gaussian
criterion \cite{EntQu}, it can be seen as as a local uncertainty
relation \cite{LUR}, with the difference that it involves a bound,
which is given by expectation values of operators, rather than
uncertainty limits represented by fixed numbers.
The inequality is optimal for the state under consideration.
Moreover it is robust against the sources of noise to be expected in
a realistic setup and provides a possibility for successful
detection of entanglement even for highly mixed states.
The inseparability criterion is presented and proven in section
\ref{Entanglement Criterion}. In this section we also consider the
influence of particle losses and find that for symmetric particle
losses in Alice's and Bob's systems, loss of a fraction up to $2/3$
of all particles can be tolerated. In section \ref{Entanglement in
superradiant scattering from Bose-Einstein condensates} we
illustrate the application of the inseparability criterion by means
of two specific examples in the context of superradiant light
scattering from Bose-Einstein condensates. More precisely, we
consider the creation of entanglement between atoms and light in
superradiant Raman scattering and describe a scheme in which
entanglement between a moving condensate and a condensate at rest is
created and purified.
\section{Inseparability criterion based on particle number measurements}\label{Entanglement Criterion}

As outlined above, the proposed entanglement criterion is optimal
for the state $|TMSS\rangle \otimes |TMSS\rangle$. In the Fock basis
this quantum state is given by
\begin{eqnarray}
|\Psi_{1}\rangle= (1-\Lambda^2)\sum_{n=0}^\infty \Lambda^n
|n\rangle|n\rangle\otimes\sum_{m=0}^\infty \Lambda^m
|m\rangle|m\rangle\label{EPRstate},
\end{eqnarray}
where $\Lambda=\tanh(r)$ and $r \in \mathbb{C}$ is the squeezing
parameter. The second and fourth ket refer to Alice's system, which
is described by two modes with creation operators $a_{+}^{\dag}$ and
$a_{-}^{\dag}$. Likewise, the first and third ket refer to Bob's
system, which is described by two modes with creation operators
$b_{+}^{\dag}$ and $b_{-}^{\dag}$.
Using this notation, we change to the Schwinger representation and
define Stokes operators $S_{x}$, $S_{y}$ and $S_{z}$ for Alice's
system and $J_{x}$, $J_{y}$ and $J_{z}$ for Bob's. $\mathbf{S}$ is
given by
\begin{eqnarray}\label{Stokesoperators}
S_{x}&=&(n_{A,x}-n_{A,y})/2,\nonumber\\
S_{y}&=&(n_{A,+45}-n_{A,-45})/2,\\
S_{z}&=&(n_{A,+}-n_{A,-})/2\nonumber,
\end{eqnarray}
and $\mathbf{J}$ is defined by analogous expressions. The number
operators $n$ carry subscripts for Alice's/Bob's system($A$/$B$) and
for the three different bases ($x/y$, $+45/-45$, $+/-$), where
$n_{A_,\pm}=a_{\pm}^{\dag}a_{\pm}$ and
$n_{B_,\pm}=b_{\pm}^{\dag}b_{\pm}$.
\subsection{Entanglement criterion}
The main characteristic feature of the two mode squeezed state
$|\Psi_{1}\rangle$ is the correlation of particle numbers in Alice's
and Bob's system. In contrast, any separable state satisfies a lower
bound for the difference in particle numbers in different mutually
independent bases. More precisely, any $2\times2$ - mode bipartite
separable state, $\rho=\sum_{i}p_{i}\rho_{i}$,
$\rho_{i}=\rho_{i}^A\otimes \rho_{i}^B$ (where $p_{i}\geqslant0$ and
$\sum_{i}p_{i}=1$), satisfies the inequality
\begin{eqnarray}\label{Criterion}
&&\langle({J_{x}-S_{x}})^{2}\rangle_{\rho}+\langle({J_{y}+S_{y}})^{2}\rangle_{\rho}+\langle({J_{z}-S_{z}})^{2}\rangle_{\rho}\nonumber\\
&&\geq \left(\langle n_{A}\rangle_{\rho}+\langle
n_{B}\rangle_{\rho}\right)/2.
\end{eqnarray}
In the following, we prove this entanglement criterion. For any
$\rho_{i}$, the left side of inequality (\ref{Criterion}) equals
\begin{eqnarray}\label{leftside}
&&\langle J_{x}^{2}+J_{y}^{2}+J_{z}^{2}\rangle_{\rho_{i}} + \langle
S_{x}^{2}+S_{y}^{2}+S_{z}^{2}\rangle_{\rho_{i}}\nonumber\\
 &&-2 \langle
J_{x}S_{x}-J_{y}S_{y}+J_{z}S_{z}\rangle_{\rho_{i}}.
\end{eqnarray}
$\langle
J_{x}^{2}+J_{y}^{2}+J_{z}^{2}\rangle_{\rho_{i}}=\langle(n_{B}/2)\left((n_{B}/2)+1\right)\rangle_{\rho_{i}}$,
where $n_{B}=n_{B,+}+n_{B,-}$, and an analogous equality holds for
the second term, as can be inferred from definition
(\ref{Stokesoperators}).
Since $\rho_{i}$ is assumed to be separable, the third term in
expression (\ref{leftside}) can be expressed as a product of two
expectation values
\begin{eqnarray*}
-2\langle \mathbf{J}\ \mathbf{\tilde{S}}\rangle_{\rho_{i}}=-2\langle
\mathbf{J}\rangle_{\rho_{i}}\langle\mathbf{\tilde{S}}\rangle_{\rho_{i}}
\end{eqnarray*}
where $\mathbf{\tilde{S}}=\left(
                      \begin{array}{ccc}
                        S_{x} & -S_{y} & S_{z} \\
                      \end{array}
                    \right)^T$.
Using $\langle\mathbf{J}\rangle_{\rho_{i}} \leq\langle
n_{B}/2\rangle_{\rho_{i}}$ and $\langle\mathbf{\tilde{S}}\rangle
_{\rho_{i}}\leq\langle n_{A}/2\rangle_{\rho_{i}}$, we obtain that
expression (\ref{leftside}) is greater than or equal to
\begin{eqnarray*}
\!&&\!\!\!\!\!\!\!\!\left\langle\frac{n_{B}}{2}\left(\frac{n_{B}}{2}+1\right)\right\rangle_{\!\!\rho_{i}}\!\!+\left\langle\frac{n_{A}}{2}\left(\frac{n_{A}}{2}+1\right)\right\rangle_{\!\!\rho_{i}}
-2\left\langle\frac{n_{B}}{2}\right\rangle_{\!\!\rho_{i}}\!\!\left\langle\frac{n_{A}}{2}\right\rangle_{\!\!\rho_{i}}.
\end{eqnarray*}
Since $\langle n_{A}/2\rangle_{\rho_{i}}\langle
n_{B}/2\rangle_{\rho_{i}}=\langle n_{A}n_{B}/4\rangle_{\rho_{i}}$
for product states, we can reexpress this equation by
\begin{eqnarray*}
\left\langle\left(\frac{n_{B}}{2}-\frac{n_{A}}{2}\right)^{2}\right\rangle_{\!\!\rho_{i}}+\left\langle\frac{n_{B}}{2}+\frac{n_{A}}{2}\right\rangle_{\!\!\rho_{i}}\geq\left\langle\frac{n_{B}}{2}+\frac{n_{A}}{2}\right\rangle_{\!\!\rho_{i}}.
\end{eqnarray*}
As this is true for every $\rho_{i}$, the average $\langle\left(
n_{A}+n_{B}\right)/2\rangle_{\rho}$ is a lower bound for the mixture
$\rho=\sum_{i}p_{i}\rho_{i}$.\\

This limitation imposed on convex mixtures of product states can be
overcome if entanglement is involved. In particular,
\begin{eqnarray*}
&&\langle(J_{x}-S_{x})^{2}\rangle_{|\Psi_{1}\rangle\langle\Psi_{1}|}+\langle(J_{y}+S_{y})^{2}\rangle_{|\Psi_{1}\rangle\langle\Psi_{1}|}\\
&&+\langle(J_{z}-S_{z})^{2}\rangle_{|\Psi_{1}\rangle\langle\Psi_{1}|}=0,
\end{eqnarray*}
as the two mode squeezed state $|\Psi_{1}\rangle$ is a simultaneous
eigenstate of $(J_{x}-S_{x})$, $(J_{y}+S_{y})$ and $(J_{z}-S_{z})$
with common eigenvalue $0$.
\subsection{Implications of particle losses}
In this subsection, we analyze how particle losses impair the
performance of the presented entanglement criterion. The influence
of particle losses is modeled by a beam splitter transformation.
Creation operators for atoms and light transform according to
\begin{eqnarray*}
a_{\pm}^{\dag}&\mapsto&
\sqrt{1-r_{A}}a_{\pm}^{\dag}-i\sqrt{r_{A}}\ v_{A\pm}^{\dag},\\
b_{\pm}^{\dag}&\mapsto& \sqrt{1-r_{B}}b_{\pm}^{\dag}-i\sqrt{r_{B}}\
v_{B\pm}^{\dag},
\end{eqnarray*}
corresponding to a beamsplitter with reflectivity $r_{A}$ for
Alice's system and $r_{B}$ for Bob's system. The quantum noise
operators $v_{A\pm}^{\dag}$, and $v_{B\pm}^{\dag}$ obey canonical
commutation relations for each mode and are mutually independent.
For the target state $|\Psi_{1}\rangle$, the left side of condition
(\ref{Criterion}) is transformed into
\begin{eqnarray}\label{Noise}
\frac{3}{2}\left((r_{A}-r_{B})^{2}(\Delta
n)^2+\left(r_{A}(1-r_{A})+r_{B}(1-r_{B})\right)\left\langle
n\right\rangle \right)\!,
\end{eqnarray}
where $\langle n\rangle=\sinh^2(r)$ and $(\Delta n)^2=2\sinh^4(r)$
is the variance of the particle number\footnote{$\langle n\rangle$
denotes here the average number of particles in any of the
considered modes $\langle n\rangle=\langle
n_{A,+}\rangle_{|\Psi_{1}\rangle\langle\Psi_{1}|}=\langle
n_{A,-}\rangle_{|\Psi_{1}\rangle\langle\Psi_{1}|}=\langle
n_{B,+}\rangle_{|\Psi_{1}\rangle\langle\Psi_{1}|}=\langle
n_{B,-}\rangle_{|\Psi_{1}\rangle\langle\Psi_{1}|}$.}. By applying
the same beamsplitter transformation to the right side of inequality
(\ref{Criterion}) one obtains
\begin{eqnarray*}
(1-r_{A})\langle{n}\rangle+(1-r_{B})\langle{n}\rangle.
\end{eqnarray*}
In case of symmetric losses $r_{A}=r_{B}=r$, successful entanglement
verification requires therefore $r<2/3$. Limitations imposed by
particle losses which are different for Alice's and Bob's system are
more restrictive, as they impair directly the symmetry-property to
which the criterion is tailored to.
For large particle numbers, the first term in expression
(\ref{Noise}) is likely to hinder the detection of entanglement.
This problem can be resolved by introducing gain factors $g_{A}$ and
$g_{B}$ for Alice and Bob, which characterize the amplification of
measured signals
\begin{eqnarray*}
a_{\pm}^{\dag}&\mapsto&
\sqrt{g_{A}(1-r_{A})}a_{\pm}^{\dag}-i\sqrt{r_{A}}\ v_{A\pm}^{\dag},\\
b_{\pm}^{\dag}&\mapsto&
\sqrt{g_{B}(1-r_{B})}b_{\pm}^{\dag}-i\sqrt{r_{B}}\ v_{B\pm}^{\dag}.
\end{eqnarray*}
In this situation, we obtain the result
\begin{eqnarray*}
&&\frac{3}{2}\left(g_{B}(1-r_{B})-g_{A}(1-r_{A})\right)^{2}(\Delta
n)^2+\\
&&\frac{3}{2}\left(g_{B}r_{B}(1-r_{B})+g_{A}r_{A}(1-r_{A})\right)\langle
n\rangle,
\end{eqnarray*}
which has to be compared to
\begin{eqnarray*}
g_{B}(1-r_{B})\langle{n}\rangle+g_{A}(1-r_{A})\langle{n}\rangle.
\end{eqnarray*}
$g_{B}$ and $g_{A}$ have to be optimized according to the
experimental parameters. For large particle numbers, the quadratic
term will dominate such that $g_{A}/g_{B}=(1-r_{B})/(1-r_{A})$
renders the problem of asymmetric losses and we obtain the condition
$(r_{A}+r_{B})/2<2/3$ for successful entanglement detection.
Remarkably this threshold does not depend on the degree of
squeezing. If the probabilities for particle losses are known,
atomic and photonic signals need not to be amplified, as it is
sufficient to adjust the inequality accordingly.
\section{Entanglement in superradiant scattering from Bose-Einstein
condensates}\label{Entanglement in superradiant scattering from
Bose-Einstein condensates}

\begin{figure}[pbt]
\begin{center}
\includegraphics[width=8.5cm]{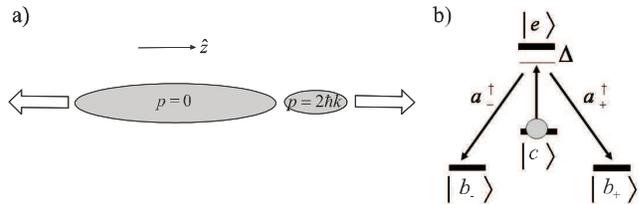}
\caption{Setup and atomic levels considered for the creation of
$N$-particle entanglement between atoms and light. a) Light
propagating along $\hat{z}$ is scattered from an elongated
condensate. Photons are scattered superradiantly into two
endfiremodes, which correspond to scattering angles of $0$ and
$\pi$, as indicated by arrows. b) Off-resonant laser light couples
to the transition $|c\rangle\rightarrow|e\rangle$, such that atoms,
which are initially prepared in $|c\rangle$ are transferred to state
or $|b_{+}\rangle$ or $|b_{-}\rangle$ via a Raman process and emit a
photon in $+$ or $-$ polarization respectively.}\label{Setup}
\end{center}
\end{figure}

As explained above, the presented entanglement criterion can be
applied in many different settings. We describe here the
verification of entanglement produced in superradiant scattering of
laser light from a Bose Einstein condensate. More specifically, we
consider a Bose Einstein condensate, which is elongated along
$\hat{z}$ and excited by a laser field propagating along the same
direction. The scattering interaction is assumed to be well within
the superradiant regime, such that light is predominantly emitted
along two endfiremodes \cite{MM99}, along $\hat{z}$ and $-\hat{z}$ ,
as shown in figure \ref{Setup}a. Atoms scattering photons along
$-\hat{z}$ acquire a momentum kick of $2\hbar k$, where $k$ is the
wave vector of the incoming light field. These atoms get spatially
separated from the BEC and form a new moving condensate. To begin
with, only the endfiremode consisting of photons scattered along
$-\hat{z}$ and atoms traveling along the $\hat{z}$ direction are
considered.
%
%
%
%
%
\subsection{Entanglement between atoms and light}

Atoms are assumed to possess an internal level structure as shown in
figure \ref{Setup}b. As was shown in \cite{CP04}, the dynamics of
the superradiant process can be described by a two mode squeezing
Hamiltonian
\begin{eqnarray*}
{H_{1}}\propto
a^{\dag}_{+}b^{\dag}_{+}+a^{\dag}_{-}b^{\dag}_{-}+H.C.\ ,
\end{eqnarray*}
where the creation operators $a^{\dag}_{+}$ and $a^{\dag}_{-}$
denote the scattered light fields in plus and minus polarization,
while $b^{\dag}_{+}$ and $b^{\dag}_{-}$ are the creation operators
for the respective atomic states. This leads to the generation of
the two mode squeezed state $|\Psi_{1}\rangle$.
Atom and photon numbers are correlated for each polarization and
inseparability of the produced quantum state can be verified
according to criterion (\ref{Criterion}) by identifying Alice with
the light field and Bob with the atomic system.
In the considered physical setting, various sources of noise may
impair the verification of entanglement. Apart from particle losses,
which have been discussed in the previous section, undesired atomic
transitions can degrade the reliability of the proposed criterion,
for instance when atoms are scattered into states other than
$|b_{-}\rangle$ and $|b_{+}\rangle$, while emitting photons in $+$
or $-$ polarization. These processes can be avoided by a suitable
choice of atomic levels. As an example we consider typical Alkali
atoms used in BEC experiments, $^{87}\text{Rb}$ and
$^{23}\text{Na}$, which have nuclear spin $3/2$. By preparing the
atomic sample in the $F=1, \ m_{F}=0$ ground state and inducing
transitions to a manifold with $F'=0,\ m_{F'}=0$, atoms can only be
scattered back to the $F=1$ groundstate manifold, occupying the
states $|F=1, m_{F}=-1\rangle\equiv|b_{-}\rangle$ and $|F=1,
m_{F}=+1\rangle\equiv|b_{+}\rangle$, while transitions to other
states are forbidden due to the selection rule $\Delta(F)=1$.
Unintentional transitions may also be mediated by interatomic
collisions. The effect of transitions from $|b_{+}\rangle$ or
$|b_{-}\rangle$ to other states is already included in consideration
of particle losses above, but the creation of a pair of atoms in
these two states without the production of the corresponding photon
pair has to be avoided. This can be done by applying electromagnetic
fields imposing Stark shifts on the internal levels such that such a
transition is prohibited by energy conservation.

The measurement of Stokes operators of light required for the
verification of the entanglement can be performed in a standard
fashion \cite{AMOreview}. The measurement of the atomic collective
spin projection $J_{z}$ can be done by counting atoms in the final
states $+,- $ with resonant absorptive imaging. The measurement of
the projections $J_{x,y}$ can be performed by applying suitable
radio-frequency $\pi/2$ pulses to the final atomic states and then
doing absorptive imaging and atom counting.
\subsection{Entanglement between two condensates}
The correlations between atoms and light, that are generated in the
process described above, can be used to create entanglement between
two condensates. To this end,  both endfire modes are considered.
The full Hamiltonian is given by
\begin{eqnarray*}
H_{2}\propto
a^{\dag}_{+I}b^{\dag}_{+I}+a^{\dag}_{-I}b^{\dag}_{-I}+a^{\dag}_{+II}b^{\dag}_{+II}+a^{\dag}_{-II}b^{\dag}_{-II}+H.C.\
,
\end{eqnarray*}
where the subscript $I$ refers to the backward scattered light and
and the moving condensate, while the subscript $II$ refers to
forward scattered light and the condensate at rest.
After the interaction, the backward scattered light field and the
moving condensate are in a two mode squeezed state as well as the
light field scattered in forward direction and the part of the
condensate at rest, which is transferred to state $|b_{+}\rangle$ or
$|b_{-}\rangle$,
\begin{eqnarray*}
|\Psi_{2}\rangle&=&
(1-\Lambda)^4\sum_{n=0}^{\infty}\Lambda^{n}|n\rangle|n\rangle\otimes
\sum_{j=0}^{\infty}\Lambda^{j}|j\rangle|j\rangle\\
&& \otimes \sum_{m=0}^{\infty}\Lambda^{m}|m\rangle |m\rangle\otimes
\sum_{l=0}^{\infty}\Lambda^{l}|l\rangle|l\rangle,
\end{eqnarray*}
where the first and second term refer to $I$-operators -
atom-photons pairs in plus and minus polarization respectively -
while the third and fourth term refer to $II$-operators.

The part of the resting condensate being in state $|b_{+}\rangle$ or
$|b_{-}\rangle$ can be entangled with the moving condensate by means
of entanglement swapping, i.e. by measuring EPR operators for each
polarization of light modes using homodyne detection. However, this
procedure leads to degradation of entanglement if non-maximally
entangled states are involved, and a distillation step has to be
performed afterwards to obtain a more useful resource state. It has
been shown that continuous variable entanglement cannot be distilled
using only Gaussian operations \cite{CVDistillation}
\footnote{Distillation of entanglement in two mode squeezed states
requires at least one non-Gaussian element as for example employed
in D.E. Browne, J. Eisert, S. Scheel, and M.B. Plenio, Phys. Rev. A
\textbf{67}, 062320 (2003);
T. Opartny, G. Kurizki, and D.-G.-Welsch, Phys. Rev. A. \textbf{61},
032302 (2000);
J. Fiurasek, L. Mista, and R. Filip, Phys. Rev. A \textbf{67},
022304 (2003);
L.-M. Duan, G. Giedke, J.I. Cirac and P. Zoller, Phys. Rev. Lett,
\textbf{84}, 4002 (2000);
L.-M. Duan, M. Lukin, J.I. Cirac, and P. Zoller, Nature
\textbf{414}, 413 (2001)}.
In the following, we describe therefore a scheme which relies on
photon counting rather than Gaussian measurements and exhibits
probabilistic entanglement distillation.
More specifically, the moving and resting condensates are entangled
by combining the forward and backward scattered light modes at a
beamsplitter and measuring the photon numbers at both output ports.
This has to be done for each polarization separately. We explain the
scheme for the $+$ polarized part of the light field. Analogous
expressions hold for the $-$ polarized part.\\

%
%
\begin{figure}[pbt]
\begin{center}
\includegraphics[width=6.5cm]{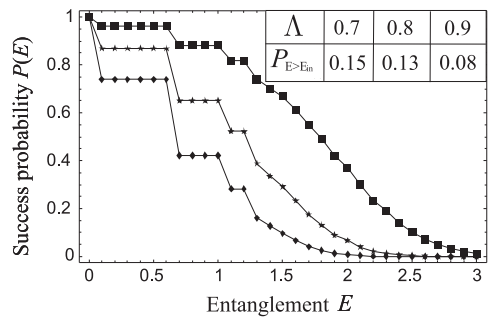}
\caption{Success probability versus entanglement which can be
produced by means of the proposed scheme for different values of
$\Lambda$. Diamonds: $\Lambda=0.7$, stars: $\Lambda=0.8$, squares:
$\Lambda=0.9$. The inset shows the probability of obtaining at least
as much entanglement as was present in the input state
$|\Psi_{1}\rangle$.}\label{Distillation}
\end{center}
\end{figure}
%
%
By applying a balanced beamsplitter transformation
$a^{in}_{+I}\rightarrow(a^{out}_{+I}+a^{out}_{+II})/\sqrt{2}$,
$a^{in}_{+II}\rightarrow(a^{out}_{+I}-a^{out}_{+II})/\sqrt{2}$ ,
where $a^{in}_{+I}$/$a^{in}_{+II}$ and
$a^{out}_{+I}$/$a^{out}_{+II}$ denote annihilation operators of the
light fields at the input and output ports of the beamsplitter
respectively, to state (\ref{EPRstate}), we obtain
\begin{eqnarray*}
|\Psi_{1}^{BS}\rangle&=&(1-\Lambda^2)\sum_{n,m=0}^{\infty}\Lambda^{n+m}\frac{1}{\sqrt{n!m!}}2^{\frac{-(n+m)}{2}}\sum_{i=0}^{n}\sum_{j=0}^{m}\\
&& \left(
  \begin{array}{c}
    n \\
    i \\
  \end{array}
\right) \left(
  \begin{array}{c}
    m \\
    j \\
  \end{array}
\right)(-1)^j \sqrt{(i+j)!(n+m-i-j)!}\\
&&|n,i+j\rangle|m,n+m-i-j\rangle.
\end{eqnarray*}
The probability of detecting $N_{I }$ photons at the first, and
$N_{II}$ photons at the second output port of the beamsplitter is
$P_{N_{I},N_{II}}=(1-\Lambda^2)\Lambda^{N_{I}+N_{II}}$. Such an
event results in the quantum state
\begin{eqnarray*}
|\Psi_{N_{I},N_{II}}\rangle&=&\sum_{n=0}^{N}k_{N_{I},N_{II}}(n)\
|n,N-n\rangle,
\end{eqnarray*}
where $N=N_{I}+N_{II}$ denotes the total number of detected photons.
The coefficients $k_{N_{I},N_{II}}(n)$ are given by
\begin{eqnarray*}
k_{N_{I},N_{II}}(n)&=&2^{\frac{-N}{2}}(-1)^{N_{I}}\sqrt{\frac{N_{II}!}{N_{I}!}}\sqrt{\frac{(N-n)!}{n!}}\frac{1}{(N_{II}-n)!}\\
&&_{2}F_{1}(-n,-N_{I},N_{II}-n+1;-1)
\end{eqnarray*}
where $_{2}F_{1}(a,b,c;z)/(c-1)!$ is the regularized hypergeometric
function. This state describes now pairs of atoms in $|b_{+}\rangle$
in the moving condensate and at rest, which are referred to in the
first and second ket respectively. For certain measurement outcomes
this state is more entangled than $|\Psi_{1}\rangle$, such that the
state can be purified by postselection.
Figure \ref{Distillation} shows the success probability versus the
produced entanglement given by the von Neumann entropy of the
reduced density matrix of the resulting atomic state
$E(N_{I},N_{II})=\sum_{n=0}^{\infty}k^2_{N_{I},N_{II}}(n)\ln\left(k^2_{N_{I},N_{II}}(n)\right)$
for different values of $\Lambda$. Note that the initial state,
which contains infinitely many terms, is truncated by the
measurement process. In this way states with entanglement close to
the maximal degree of entanglement in the corresponding subspace can
be produced. For example, for $N_{I}=1$ and $N_{II}=0$, the
maximally entangled state
$|\Psi\rangle=(|1,0\rangle+|1,0\rangle)/\sqrt{2}$ is created.
Considering the light field in $+$ as well as in $-$ polarization,
the resulting atomic state can be detected by criterion
(\ref{Criterion}) after local transformation $n_{+II}\mapsto
N-n_{+II}$ and $n_{-II}\mapsto N'-n_{-II}$. In this case
$\mathbf{J}$ refers to atomic operators at rest and $\mathbf{S}$
describes the moving condensate.\\
\\After preparation of this manuscript, Dr. C. Simon informed us that
the entanglement criterion (\ref{Criterion}) presented in this work
coincides with inequality (4) derived in \cite{EntCrit}.
\section*{Acknowledgements}

We thank J\"org Helge M\"uller and Geza Giedke for valuable
discussions and acknowledge support from the Elite Network of
Bavaria (ENB) project QCCC, the EU projects SCALA, QAP, COVAQUIAL,
COMPAS and the DFG-Forschungsgruppe 635.

%
%

%
\end{document}